\documentclass[aps,pre,reprint,superscriptaddress,nolongbibliography]{revtex4-2}
\usepackage[italicdiff]{physics}
\usepackage{graphicx,siunitx,booktabs,xcolor,hyperref}
\hypersetup{
colorlinks,
linkcolor={blue!50!black},
citecolor={blue!50!black},
urlcolor={blue!80!black}
}
\DeclareMathOperator*{\argmax}{\arg\!\max}
\newcommand{\Nup}{N_\text{up}}
\begin{document}
\title{Non-phononic density of states of two-dimensional glasses revealed by random pinning}
\author{Kumpei Shiraishi}
\email{kumpeishiraishi@g.ecc.u-tokyo.ac.jp}
\affiliation{Graduate School of Arts and Sciences, University of Tokyo, Komaba, Tokyo 153-8902, Japan}
\author{Hideyuki Mizuno}
\affiliation{Graduate School of Arts and Sciences, University of Tokyo, Komaba, Tokyo 153-8902, Japan}
\author{Atsushi Ikeda}
\affiliation{Graduate School of Arts and Sciences, University of Tokyo, Komaba, Tokyo 153-8902, Japan}
\affiliation{Research Center for Complex Systems Biology, Universal Biology Institute, University of Tokyo, Komaba, Tokyo 153-8902, Japan}
\date{\today}
\begin{abstract}
The vibrational density of states of glasses is considerably different from that of crystals.
In particular, there exist spatially localized vibrational modes in glasses.
The density of states of these non-phononic modes has been observed to follow $g(\omega) \propto \omega^4$, where $\omega$ is the frequency.
However, in two-dimensional systems, the abundance of phonons makes it difficult to accurately determine this non-phononic density of states because they are strongly coupled to non-phononic modes and yield strong system-size and preparation-protocol dependencies.
In this article, we utilize the random pinning method to suppress phonons and disentangle their coupling with non-phononic modes and successfully calculate their density of states as $g(\omega) \propto \omega^4$.
We also study their localization properties and confirm that low-frequency non-phononic modes in pinned systems are truly localized without far-field contributions.
We finally discuss the excess density of states over the Debye value that results from the hybridization of phonons and non-phononic modes.
\end{abstract}
\maketitle

\section{Introduction}
Low-frequency vibrational states of glasses have been attracting considerable attention in recent years.
Unlike crystals~\cite{Ashcroft_Mermin}, their low-frequency vibrational modes are not described by phonons alone; there exist spatially localized vibrations.
The vibrational density of states of these non-phononic localized modes follows $g(\omega) \propto \omega^4$~\cite{Lerner_2016,Mizuno_2017}.
The localized modes are widely observed in various systems regardless of interaction potentials~\cite{Bonfanti_2020}, details of constituents~\cite{Richard_2020}, asphericity of particles~\cite{Shiraishi_2020}, and stability of configurations~\cite{Wang_low_freq_2019}.

Theoretical backgrounds of the non-phononic vibrational density of states have been studied.
Mean-field theories predict that glasses exhibit the non-Debye scaling law of $g(\omega) \propto \omega^2$ at low frequencies by both the replica theory~\cite{Franz_2015} and the effective medium theory~\cite{DeGiuli_2014}, and numerical simulations of high-dimensional packings confirm this behavior~\cite{Charbonneau_2016,Shimada2020large}.
Recently, replica theories of interacting anharmonic oscillators~\cite{Bouchbinder_2021,Folena2022Marginal} and the effective medium theory~\cite{Shimada2020Vibrational,Shimada2021Novel,Shimada2022Random} have also successfully derived the $\omega^4$ scaling of glasses.
Of these mean-field theories, the effective medium theory~\cite{DeGiuli_2014,Shimada2020Vibrational,Shimada2021Novel,Shimada2022Random} naturally deals with phonon modes together with non-phononic modes, whereas the other theories focus on non-phononic modes without particular attention on phonon modes.

Meanwhile, phonons do exist even in the amorphous solids, which strongly hybridize with the non-phononic localized modes.
In this case, the scaling of $g(\omega)$ is described by the framework of the generalized Debye model~\cite{Schirmacher_2006,Schirmacher_2007,Marruzzo_2013,Schirmacher_2014} that predicts that the exponent should be consistent with that of the Rayleigh scattering $\Gamma \propto \Omega^{d+1}$ of acoustic attenuation ($\Gamma$ is attenuation rate, $\Omega$ is propagation frequency, and $d$ is the spatial dimension).
Therefore, $g(\omega)$ is predicted to scale with $\omega^{d+1}$.
Numerical simulations of three-dimensional glasses ($d = 3$) show that the phonon attenuation rate follows $\Gamma \propto \Omega^4$~\cite{Monaco_Mossa_2009,Marruzzo_2013,Mizuno_Mossa_Barrat_2014,Mizuno_2018,Wang_sound_sttenuation_2019}.
This behavior of Rayleigh scattering has also been observed in recent experimental studies~\cite{Ruffle_2006,Monaco_Giordano_2009,Baldi_2010,Baldi_2014}.
Simulation study also reveals that the vibrational density of states follows $g(\omega) \propto \omega^4$~\cite{Mizuno_2017}.
Thus, in three-dimensional systems, acoustic attenuation and the vibrational density of states both exhibit the exponent of $d + 1 = 4$, consistent with the generalized Debye theory.

However, in two-dimensional glasses, conflicting results have been reported.
In acoustic attenuation simulations in two-dimensional glasses ($d = 2$), the Rayleigh scattering of $\Gamma \propto \Omega^3$ is indeed observed~\cite{Mizuno_2018,Saitoh_2021,Kapteijns_2021,PicaCiamarra2022network}.
In simulations of direct measurements of $g(\omega)$ in two dimensions, Mizuno \textit{et al.}\ performed the vibrational analysis of glass configurations of large system sizes and revealed that localized vibrations were too few to determine the non-phononic scaling of $g(\omega)$~\cite{Mizuno_2017}.
Afterward, Kapteijns \textit{et al.}\ reported that the $\omega^4$ scaling holds even for two-dimensional glasses by performing simulations of systems with small system sizes for a large ensemble of configurations to extract the sufficient number of modes below the first phonon frequency~\cite{Kapteijns_2018}.
However, a recent study by Wang \textit{et al.}\ reported a contradictory result of $g(\omega) \propto \omega^{3.5}$ from simulations of small systems~\cite{Wang2D2021,*Wang2022Erratum}.
More recently, Lerner and Bouchbinder suggested that the exponent depends on the glass formation protocol and system size and claimed the exponent to be 4 in the thermodynamic limit even in two dimensions~\cite{Lerner_2022}.
In contrast, a recent work by Wang \textit{et al.}\ claimed that there are no system-size effects and the exponent remains as 3.5~\cite{Wang2022Scaling}.

The above conflicting results could be due to the emergence of phonons and their coupling with the localized modes, making it difficult to accurately determine the non-phononic vibrational density of states.
Here, we utilize the random pinning method to resolve this problem.
Originally, this method is used to realize equilibrium glass states~\cite{Cammarota_2012,Kob_2013,Ozawa_2015}.
Angelani \textit{et al.}\ showed that this method can be used to suppress phononic modes and probe the non-phononic density of states~\cite{Angelani_2018}.
Recently, we revealed that low-frequency localized modes of pinned glasses are disentangled with phonons by numerical simulations of three-dimensional glasses~\cite{Shiraishi2022ideal}.
By performing vibrational analysis in two-dimensional pinned glasses, we can expect to put an end to the controversial results of the non-phononic density of states.

In this paper, we report the properties of low-frequency localized modes in two-dimensional glasses induced by randomly pinned particles.
First, we study the participation ratio of each mode and show the low-frequency modes of pinned glasses indeed have a localized character.
Second, we also study their localization properties by calculating the decay profile.
Those modes show exponentially decaying profiles, that is, they are truly localized.
Finally, we evaluate the vibrational density of states of localized modes and observe the scaling of $g(\omega) \propto \omega^4$ in two-dimensional glasses with pinned particles.
Our results elucidate the bare nature of low-frequency localized modes of glasses by obliterating harmful phononic modes using the random pinning operation.

\section{Methods}
We perform vibrational mode analyses on the randomly pinned Kob-Andersen system in two-dimensional space~\cite{Bruning_2009}, which is identical to a model studied by Wang \textit{et al.}~\cite{Wang2D2021,*Wang2022Erratum}.
We consider a system of $N$ particles with identical masses of $m$ enclosed in a square box with periodic boundary conditions.
The linear size $L$ of the box is determined by the number density of $\rho = 1.204$.
Particles A and B are mixed in a ratio of 65:35 to avoid crystallization~\cite{Bruning_2009}.
The particles interact via the Lennard-Jones potential
\begin{align}
V(r_{ij}) = \phi(r_{ij}) - \phi(r^\text{cut}_{ij}) - \phi^\prime(r^\text{cut}_{ij}) (r_{ij} - r^\text{cut}_{ij}),
\end{align}
with
\begin{align}
\phi(r_{ij}) = 4\epsilon_{ij}\bqty{\pqty{\sigma_{ij}/r_{ij}}^{12} - \pqty{\sigma_{ij}/r_{ij}}^6},
\end{align}
where $r_{ij}$ denotes the distance between interacting particles, and the cut-off distance is set to $r^\text{cut}_{ij} = 2.5\sigma_{ij}$.
The interaction parameters are chosen as follows:
$\sigma_\text{AA} = 1.0,\ \sigma_\text{AB} = 0.8,\ \sigma_\text{BB} = 0.88,\ \epsilon_\text{AA} = 1.0,\ \epsilon_\text{AB} = 1.5,\ \epsilon_\text{BB} = 0.5$.
Lengths, energies, and time are measured in units of $\sigma_\text{AA}$, $\epsilon_\text{AA}$, and $\pqty{m\sigma_\text{AA}^2/\epsilon_\text{AA}}^{1/2}$, respectively.
The Boltzmann constant $k_\text{B}$ is set to unity when measuring the temperature $T$.

To prepare the randomly pinned system, we first run molecular dynamics simulations in the \textit{NVT} ensemble to equilibrate the system in the normal liquid state at $T = 5.0$ for the time of $t = 2.0 \times 10^2$, which is sufficiently longer than the structural relaxation time.
After the equilibration, we randomly choose particles and freeze their positions.
The FIRE algorithm~\cite{Guenole_2020} is applied to the system to minimize energy (stop condition is $\max_i F_i < 3.0 \times 10^{-10}$), which produces the glass-solid state at zero temperature, $T = 0$.
We denote the fraction of pinned particles as $c$ ($0 \leq c \leq 1$) and the number of unpinned (vibrating) particles as $\Nup = (1 - c)N$.
Then, we perform the vibrational mode analysis on the randomly pinned system and obtain the eigenvalues $\lambda_k$ and eigenvectors $\vb*{e}_k = \pqty{\vb*{e}_k^1, \vb*{e}_k^2, \dots, \vb*{e}_k^{\Nup}}$, where $k = 1, 2, \dots, 2\Nup$~\cite{MizunoIkeda2022}.
Note that two zero-frequency modes corresponding to global translations do not appear in pinned systems because the existence of pinned particles breaks translational invariance~\cite{Angelani_2018,Shiraishi2022ideal}.
For details of the random pinning procedure and vibrational mode analysis, please refer to Ref.~\onlinecite{Shiraishi2022ideal}.

\section{Results}
\subsection{Participation ratio}
\begin{figure}
\centering
\includegraphics[width=\linewidth]{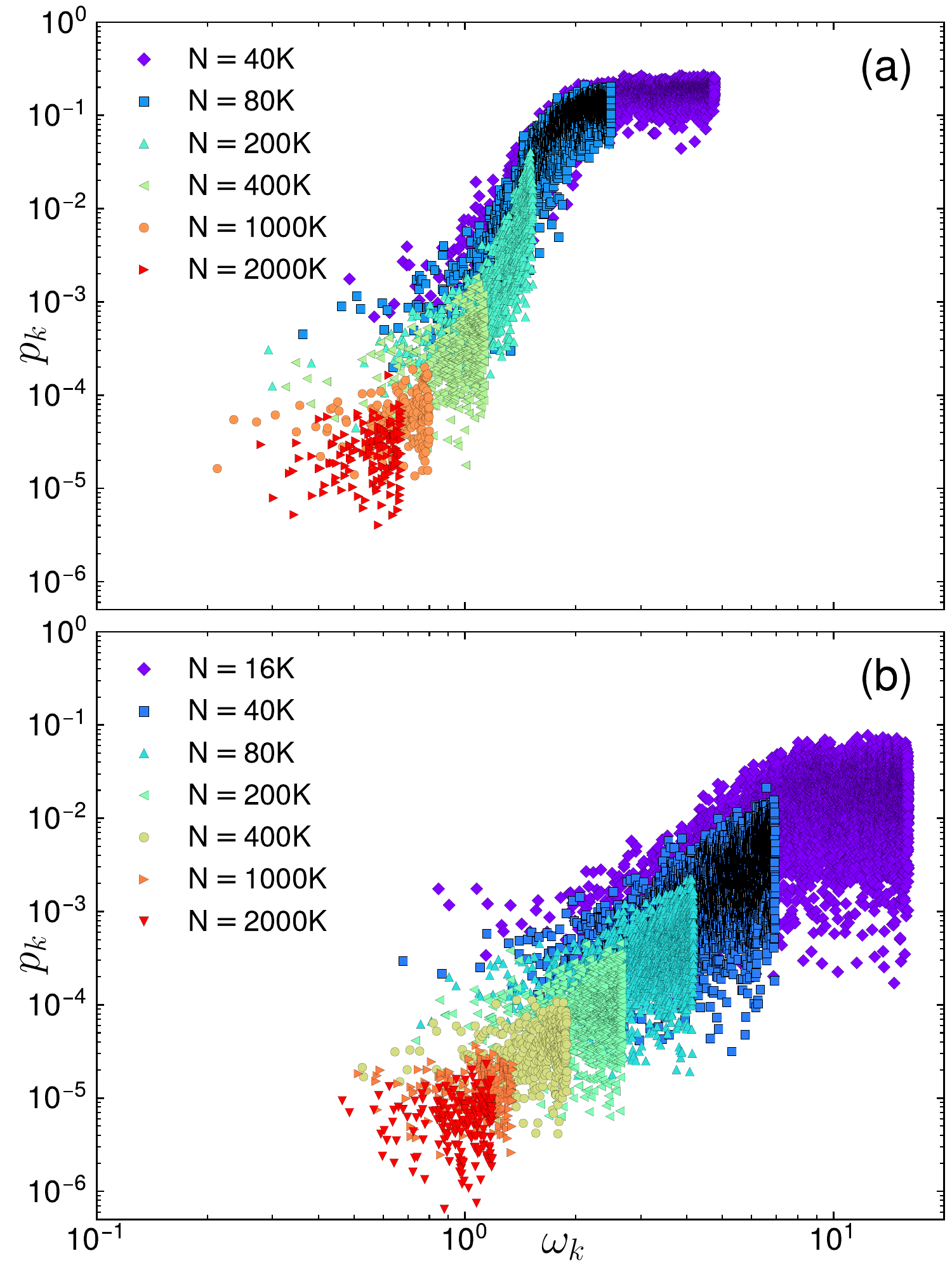}
\caption{Participation ratio $p_k$ versus mode frequencies $\omega_k$.
The figures show the data of the lowest-frequency region.
The fractions $c$ of pinned particles are (a) 0.03 and (b) 0.20.}
\label{fig:PR}
\end{figure}

First, we study the participation ratio
\begin{align}
 p_k = \frac{1}{\Nup\sum_{i=1}^{\Nup} \abs{\vb*{e}_k^i}^4},
\end{align}
which quantifies the fraction of particles that participate in the mode $k$~\cite{Schober_1991,mazzacurati_1996}.
Figure~\ref{fig:PR} shows $p_k$ versus eigenfrequencies $\omega_k = \sqrt{\lambda_k}$ for $c = 0.03$ and $c = 0.20$.
The number of particles in the systems ranges from $N = \num{16000}$ to $N = \num{2000000}$.

As we can easily recognize from Fig.~\ref{fig:PR}, pinned systems have numerous localized modes with low $p_k$ in the low-frequency region.
This result is strikingly different from the unpinned system, where most low-frequency modes are spatially extended phonons and localized modes are hard to observe in two dimensions~\cite{Mizuno_2017}.
When pinned particles are introduced, these phonon modes are suppressed because translational invariance is violated, and low-frequency localized modes emerge, like in three-dimensional glasses~\cite{Angelani_2018,Shiraishi2022ideal}.
Comparing the cases of $c = 0.03$ and $0.20$ shows that $p_k$ is lower in $c = 0.20$.
In particular, when $c = 0.20$, there are modes whose participation ratio is near $p_k = 1/\Nup$, indicating that only one particle out of $\Nup$ particles vibrates in the mode $k$.

As mentioned in the Introduction, the difficulty of studying low-frequency localized modes in two-dimensional glasses originates from the abundance of low-frequency phonons~\cite{Lerner_2022}.
As demonstrated in Fig.~\ref{fig:PR}, phonon modes are well-suppressed in two-dimensional pinned glasses, and only low-frequency localized modes remain.
Our data of $p_k$ clearly shows that random pinning excludes phonons and resolves this difficulty for analysis on non-phononic modes in two-dimensional glasses.

\subsection{Decay profile}
\begin{figure}
\centering
\includegraphics[width=\linewidth]{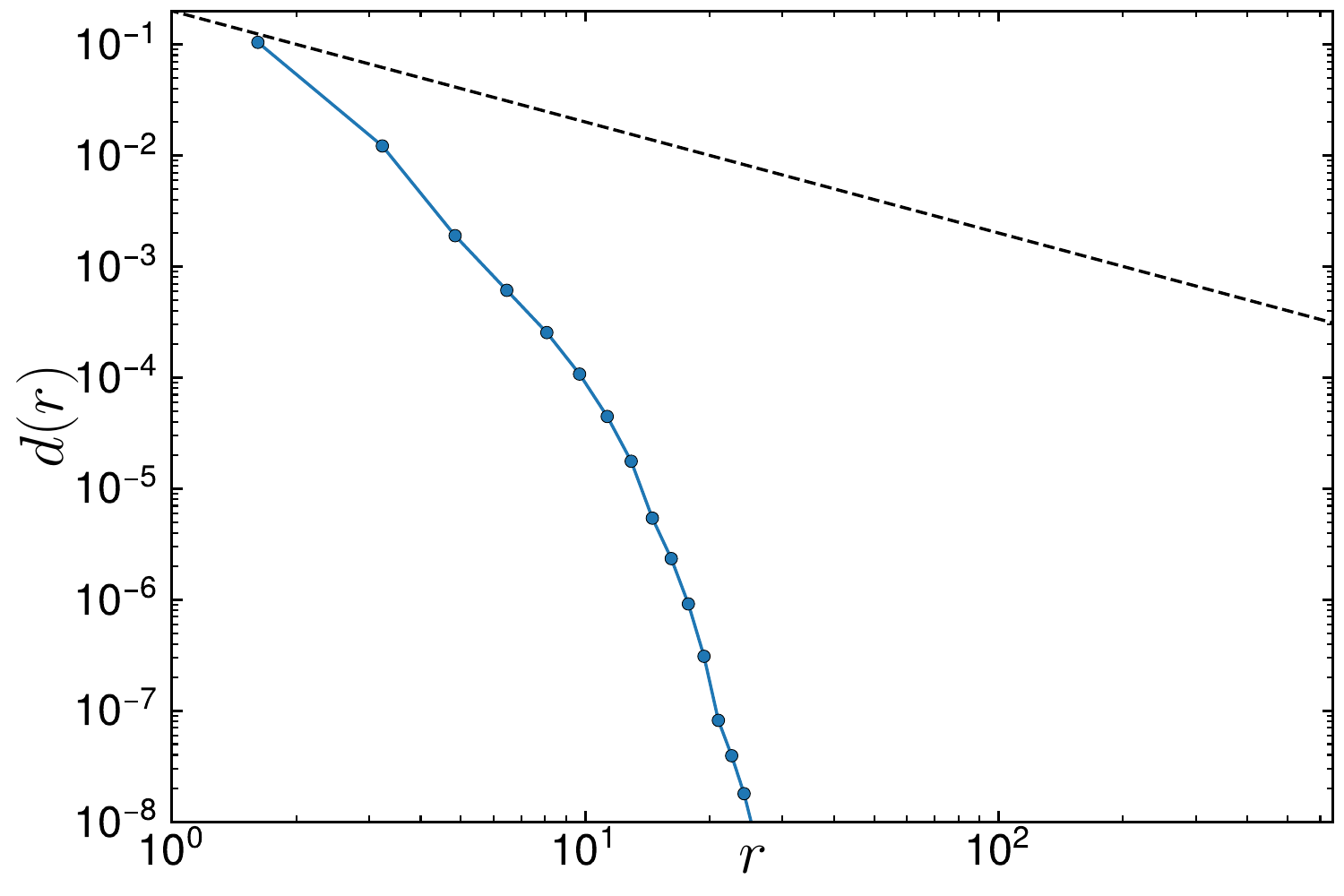}
\caption{Decay profile $d(r)$ of a low-frequency mode of the system with $N = \num{2000000}$ and $c = 0.20$.
The mode has the eigenfrequency of $\omega_k = 0.5968$ and participation ratio of $p_k = 1.9 \times 10^{-6}$.
The dashed line indicates the power-law behavior of $d(r) \propto r^{-1}$.}
\label{fig:decay}
\end{figure}

Next, we scrutinize the spatial structure of a low-frequency mode by calculating the decay profile $d(r)$ as in Ref.~\onlinecite{Lerner_2016}, which is defined as
\begin{align}
d(r) = \frac{\abs{\vb*{e}_k^i}}{\max_i \abs{\vb*{e}_k^i}}.
\end{align}
When calculating $d(r)$, we take the median of each contribution $\abs{\vb*{e}_k^i}$ from particles inside a shell with radius $r$ from the most vibrating particle $i_\text{max} = \argmax_i \abs{\vb*{e}_k^i}$.
Figure~\ref{fig:decay} presents the decay profile $d(r)$ of a low-frequency mode of a configuration with $N = \num{2000000}$ and $c = 0.20$ ($\Nup = \num{1600000}$).

As in Fig.~\ref{fig:decay}, the decay profile of pinned glasses deviates from the power-law behavior of $d(r) \propto r^{-1}$~\cite{Kapteijns_2018}.
Instead, $d(r)$ shows an exponential decay, consistent with the behavior in three-dimensional pinned glasses~\cite{Shiraishi2022ideal}.
This result indicates that the spatial structures of low-frequency localized modes are significantly different from those of unpinned glasses~\cite{Lerner_2016,Kapteijns_2018,Shimada_spatial_2018}.
The power-law decay of $d(r) \propto r^{-1}$, which is missing in Fig.~\ref{fig:decay}, is a consequence of the absence of hybridization with phonons~\cite{Lerner_2016}.
Therefore, we conclude that the random pinning method prevents non-phononic localized modes from hybridizing with phonon modes, as in the three-dimensional system~\cite{Shiraishi2022ideal}.

\subsection{Vibrational density of states}
\begin{figure}
\centering
\includegraphics[width=\linewidth]{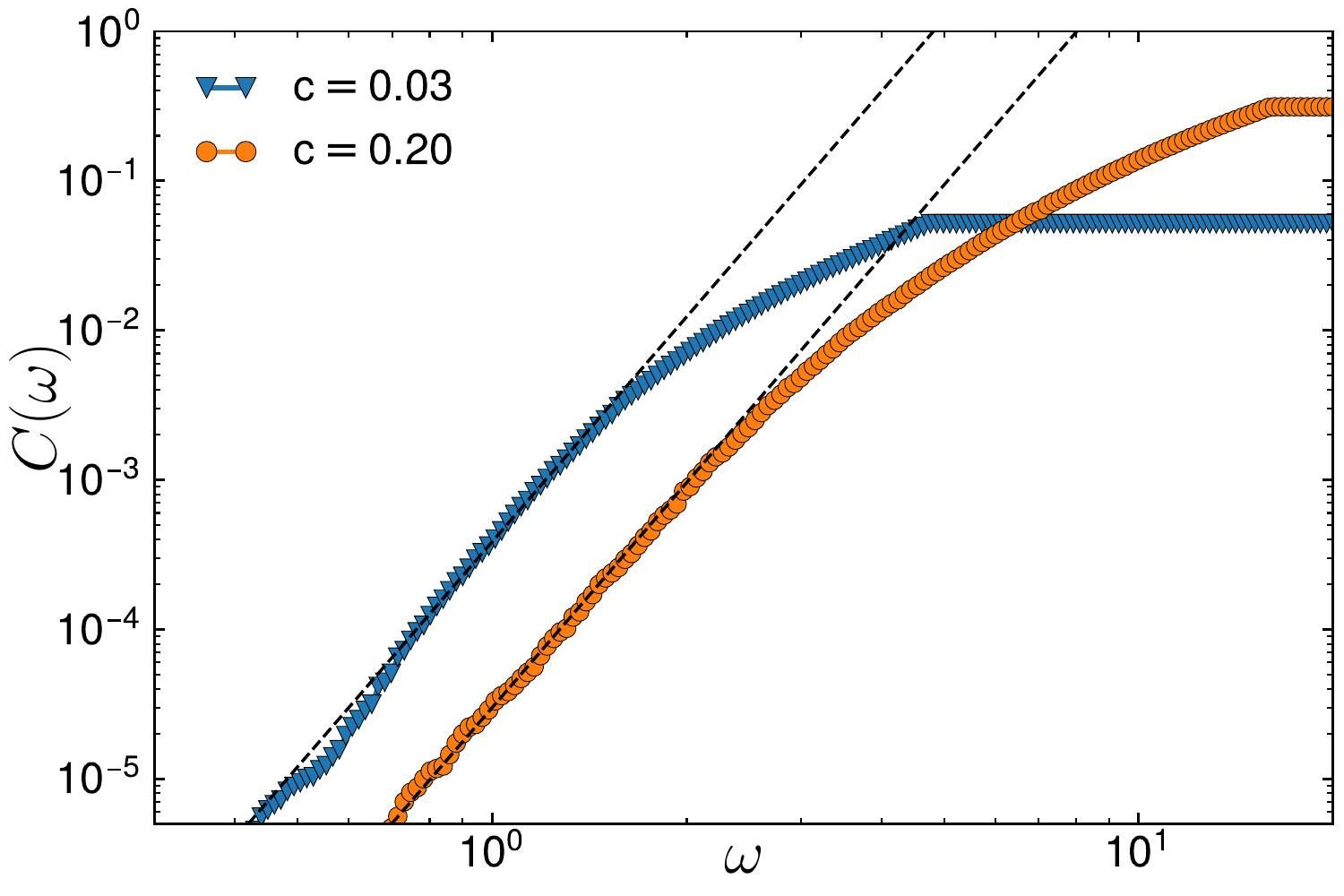}
\caption{Cumulative density of states $C(\omega)$ of systems with $c = 0.03$ and $c = 0.20$.
The dashed lines indicate $C(\omega) \propto \omega^5$.}
\label{fig:cdf}
\end{figure}

Finally, we study the vibrational density of states in the low-frequency regime of randomly pinned two-dimensional glasses.
The vibrational density of states is calculated as
\begin{align}
 g(\omega) = \frac{1}{N_\text{mode}} \sum_k \delta (\omega - \omega_k),
\end{align}
where $N_\text{mode} = 2\Nup$ is the number of all eigenmodes and $\delta(x)$ is the Dirac delta function.
However, the value of $g(\omega)$ is sensitive to the binning setups used for the calculation.
To determine the density of states without the arbitrariness of binning, we present the cumulative density of states:
\begin{align}
 C(\omega) = \int_0^\omega g(\omega^\prime) \dd{\omega^\prime}.
\end{align}
Figure~\ref{fig:cdf} presents $C(\omega)$ for $c = 0.03$ and $c = 0.20$.
When generating Fig.~\ref{fig:cdf}, we averaged $C(\omega)$ of different system sizes presented in Fig.~\ref{fig:PR}.
The results from these different system sizes provide information for the very low-frequency regime~\cite{Mizuno_2017}.
We recall that each of these systems has the fraction $c$ of pinned particles; therefore, the number of vibrating particles $\Nup$ is smaller than $N$.

As shown in Fig.~\ref{fig:cdf}, $C(\omega)$ obeys $\omega^5$ scaling, that is, the vibrational density of states obeys $g(\omega) \propto \omega^4$ in the low-frequency regime, which is the main result of this work.
This behavior is consistent with various reports in three-dimensional glasses~\cite{Lerner_2016,Mizuno_2017,Angelani_2018,Shiraishi2022ideal}.
Our result is also consistent with a report by Kapteijns \textit{et al.}~\cite{Kapteijns_2018} who studied two-dimensional unpinned glasses of small systems.

Here, we emphasize that the random pinning method suppresses phononic modes and enables us to directly probe the non-phononic density of states without generating a large ensemble of small systems.
Furthermore, because the low-frequency modes of pinned glasses do not hybridize with phonons, the investigation of $g(\omega)$ with random pinning is free of finite-size effects~\cite{Lerner_PRE_2020,Lerner_2022} or glass-formation-protocol dependence~\cite{Lerner_2022} appearing in $g(\omega)$.

\section{Discussions}
In summary, we report the properties of low-frequency vibrations of two-dimensional glasses with randomly pinned particles.
While there exist a large number of phonon modes in two-dimensional glasses, which cause hybridization with non-phononic modes, the random pinning operation can well suppress phonon modes to disentangle their hybridization.
We confirm the disentanglement numerically by observing the participation ratio and decay profile and conclude that non-phononic modes are truly localized modes that are not coupled to phonon modes.
Therefore, we can easily intrude on the non-phononic density of states of localized modes at low frequencies.
Then, our main result is that the cumulative vibrational density of states of non-phononic modes obeys $C(\omega) \propto \omega^5$, that is, the vibrational density of states follows $g(\omega) \propto \omega^4$ in two-dimensional glasses.
This result provides a sound basis for the controversial vibrational density of states of two-dimensional glasses and could resolve the conflicting reports of the exponent~\cite{Kapteijns_2018,Wang2D2021,*Wang2022Erratum,Lerner_2022,Wang2022Scaling}.
Our work also demonstrates the benefit of the random pinning method, not only for the glass transition studies but also for the material properties of amorphous solids.

\begin{table}
\centering
\caption{Dimensional dependence of the exponent of the excess density of states $g(\omega) \propto \omega^\beta$ (with phonons) with corresponding values of the non-phononic density of states (without phonons).
Note that $\beta$ has not yet been measured for the excess density of states (with phonons) and $d \geq 4$; however, we might expect $\beta = 4$ (see the main text).}
\label{table:exponent}
\begin{tabular}{cll}
\toprule
Dimension  & With phonons                   & Without phonons \\
\midrule
$d = 2$    & $\beta = 3$~\cite{Mizuno_2018} & $\beta = 4$ (this paper) \\
$d = 3$    & $\beta = 4$~\cite{Mizuno_2018} & $\beta = 4$~\cite{Shiraishi2022ideal} \\
$d \geq 4$ & $\beta = 4$ (expected)         & $\beta = 4$~\cite{Kapteijns_2018} \\
\bottomrule
\end{tabular}
\end{table}

Our present analysis of randomly pinned two-dimensional glasses reveals the vibrational density of states of non-phononic modes that are completely free from hybridization with phonons.
On the other hand, in the following, we discuss the ``excess'' density of states over the Debye value in a situation where abundant phonon modes exist and hybridize with non-phononic modes.
Here, we use the phrase ``excess density of states'' because we generally cannot distinguish non-phonon modes from phonon modes when they are strongly hybridized.
In this situation, we can apply the generalized Debye theory~\cite{Schirmacher_2006,Schirmacher_2007,Marruzzo_2013,Schirmacher_2014,Mizuno_2018} to measure the ``excess'' density of states, where the exponent $\beta$ of $g(\omega) \propto \omega^\beta$ is provided by the exponent $\gamma$ of the acoustic attenuation $\Gamma \propto \Omega^\gamma$.

We refer to previous studies and summarize the values of $\beta$ in Table~\ref{table:exponent} for the spatial dimensions of $d=2$, $d=3$, and $d \geq 4$.
In the table, we also present the corresponding values of $\beta$ of the non-phononic density of states without phonon modes for comparison.
For the case without phonons, where truly-localized modes are realized, the non-phononic density of states follows $g(\omega) \propto \omega^4$ for $d = 2$ and $d = 3$, as confirmed in the present ($d = 2$) and previous study~\cite{Shiraishi2022ideal} ($d = 3$) using the random pinning method.
The previous numerical work~\cite{Kapteijns_2018} also provided the value of $\beta = 4$ for $d = 2$ to $d = 4$.
In addition, the mean-field theories predicted $g(\omega) \propto \omega^4$~\cite{Bouchbinder_2021,Folena2022Marginal}, which validates the value of $\beta = 4$ for $d \geq 4$.

In contrast, for the case with phonons where non-phononic modes hybridize with phonons to become quasi-localized, the Rayleigh scattering behavior of $\Gamma \propto \Omega^\gamma = \Omega^{d+1}$ is observed in $d = 2$ and $d = 3$, leading to the value of $\beta = d+1$~\cite{Mizuno_2018}.
For the larger dimensions of $d \geq 4$, there are no numerical results so far; however, we speculate the following.
The non-phononic density of states without phonons, $g(\omega) \propto \omega^4$, shows larger orders of values than $g(\omega) \propto \omega^{d+1}$ (in the low-frequency regime) since $4 < d+1$.
Considering this, we might expect that the hybridization maintains the value of exponent $\beta = 4$ for the excess density of states.
Note that the results shown in Ref.~\onlinecite{Shimada2020large} are consistent with this expectation though their system sizes are not large enough to conclude this point.
Future studies should focus on measuring the acoustic attenuation $\Gamma$ for $d \geq 4$ and determining the scaling of $\Gamma \propto \Omega^\gamma$.
In addition, these studies should also measure $g(\omega) \propto \omega^\beta$ directly in the presence of numerous phonon modes.

Again, we emphasize that hybridization effects strongly emerge in the $d=2$ case: $g(\omega) \propto \omega^4$ (non-phononic density of states) without hybridization, whereas $g(\omega) \propto \omega^{d+1} = \omega^3$ (excess density of states) with hybridization.
Even when $d = 2$, if we resort to generating a large ensemble of small systems, the exponent of $g(\omega)$ could be measured~\cite{Kapteijns_2018,Wang2D2021,*Wang2022Erratum,Lerner_2022,Wang2022Scaling}.
However, the shortcoming of this method is that the intensity of the hybridization of modes, which appears as the distance from the first phonon level, cannot be controlled.
The hybridization with phonons causes harmful effects, such as finite-size effects~\cite{Lerner_PRE_2020,Lerner_2022} or dependence on preparation protocols~\cite{Lerner_2022}.
These effects could change the value of the exponent $\beta$ of $g(\omega)$.
Therefore, we would conclude that the exponent observed using systems with a small number of particles can fluctuate between 3 to 4, which can be the reason for the controversial results in the $d=2$ case~\cite{Kapteijns_2018,Wang2D2021,*Wang2022Erratum,Lerner_2022,Wang2022Scaling}.

\begin{acknowledgments}
This work is supported by JSPS KAKENHI (Grant Numbers 18H05225, 19H01812, 20H00128, 20H01868, 21J10021, 22K03543) and Initiative on Promotion of Supercomputing for Young or Women Researchers, Information Technology Center, the University of Tokyo.
\end{acknowledgments}

\end{document}